\documentclass[usenatbib,usegraphicx]{mn2e}
\usepackage{amssymb}
\newcommand{\halfspace}{\hspace{1pt}}
\newcommand{\sect}{\S\ref}
\newcommand{\tab}{Table~\ref}
\newcommand{\Mpc}{\mathop{\rm Mpc\,}\nolimits}
\newcommand{\kms}{\mathop{\rm km \ s^{-1}\,}\nolimits}
\newcommand{\Lya}{Ly$\alpha$}
\newcommand{\Lyb}{Ly$\beta$}
\newcommand\HI{{\hbox{H\halfspace$\rm \scriptstyle I$}}}
\newcommand\HII{{\hbox{H\halfspace$\rm \scriptstyle II$}}}
\newcommand\HeI{{\hbox{He\halfspace$\rm \scriptstyle I$}}}
\newcommand\HeII{{\hbox{He\halfspace$\rm \scriptstyle II$}}}
\newcommand\HeIII{{\hbox{He\halfspace$\rm \scriptstyle III$}}}

\newcommand\LCDM{$\Lambda$CDM}
\newcommand\lsim{~\lower.5ex\hbox{$\buildrel < \over \sim$}~}
\newcommand\gsim{~\lower.5ex\hbox{$\buildrel > \over \sim$}~}

\title[Helium reionization and the thermal proximity effect]{
       Helium reionization and the thermal proximity effect}
\author[Avery Meiksin, Eric R. Tittley and Calum K. Brown]{
        Avery Meiksin$^{1}$\thanks{E-mail:\ aam@roe.ac.uk (AM)},
        Eric R. Tittley$^{1}$ and Calum K. Brown$^{2}$\\
        $^{1}$SUPA\thanks{Scottish Universities Physics Alliance},
	Institute for Astronomy, University of Edinburgh,
        Blackford Hill, Edinburgh\ EH9\ 3HJ, UK\\
        $^{2}$Rudolf Peierls Centre for Theoretical Physics, 1 Keble Road,              Oxford\ OX1\ 3NP, UK}

\shortcites{}

\begin{document}

\date{Accepted . Received ; in original form }
\pagerange{\pageref{firstpage}--\pageref{lastpage}} \pubyear{2009}
\maketitle
\label{firstpage}

\begin{abstract}
  We examine the temperature structure of the intergalactic medium
  (IGM) surounding a hard radiation source, such as a Quasi-Stellar
  Object (QSO), as it responds to the onset of helium reionization by
  the source. We model the reionization using a radiative transfer
  (RT) code coupled to a particle-mesh (PM) $N$-body code. Neutral
  hydrogen and helium are initially ionized by a starburst spectrum,
  which is allowed to gradually evolve into a power law spectrum
  ($\propto \nu^{-0.5}$). Multiple simulations were performed with
  different times for the onset and dominance of the hard spectrum,
  with onset redshifts ranging from $z=3.5$ to 5.5. The source is
  placed in a high-density region to mimic the expected local
  environment of a QSO. Simulations with the source placed in a
  low-density environment were also performed as control cases to
  explore the role of the environment on the properties of the
  surrounding IGM. We find in both cases that the IGM temperature
  within the \HeIII\ region produced exceeds the IGM temperature
  before full helium reionization, resulting in a ``thermal proximity
  effect,'' but that the temperature in the \HeIII\ region increases
  systematically with distance from the source. With time the
  temperature relaxes with a reduced spread as a function of impact
  parameter along neighbouring lines of sight, although the trend
  continues to persist until $z=2$. Such a trend could be detected
  using the widths of intervening metal absorption systems using high
  resolution, high signal-to-noise ratio spectra. By contrast, the
  Doppler widths of \HI\ absorption features in mock spectra along
  neighbouring lines of sight show a weak trend with impact parameter
  prior to full helium reionization reflecting the behaviour of the
  underlying density and peculiar velocity fields, but they take on a
  near constant value after helium reionization, with a median value
  near $25\kms$ at $z=3$, in good agreement with observations.
\end{abstract}


\section{Introduction}
\label{sec:Intro}

An abundance of observational evidence shows that subsequent to the
Recombination Era, during which both hydrogen and helium recombined to
their neutral forms, the Universe was reionized. The best constraints
on the epoch of hydrogen reionization are derived from measurements of
the intergalactic Lyman resonance line transitions in the spectra of
high redshift Quasi-Stellar Objects (QSOs) and polarization
measurements of the Cosmic Microwave Background (CMB). The measured
optical depths of intergalactic \Lya\ and \Lyb\ absorption in high
redshift QSOs show that the Universe was reionized by $z\gsim 6$
\citep{Becker01, 2002AJ....123.1247F}. This is consistent with CMB
measurements by the {\it Wilkinson Microwave Anisotropy Probe} ({\it
  WMAP}), which place the hydrogen reionization epoch, if a sudden
process, at $z=11.0\pm1.4$ \citep{2009ApJS..180..306D}. The sources of
hydrogen reionization are currently unknown, but are strongly
suspected of being young star-forming galaxies. More speculative
possibilities include pockets of Population III stars, as may arise in
isolated star clusters \citep{1999ApJ...514..648M, CF06}, or
miniquasars \citep{Madau04}. Hydrogen reionization by QSOs is
untenable based on current estimates of the QSO luminosity function,
unless there is a sharp rise at the faint end at high redshifts
\citep{Meiksin05, 2009ApJ...692.1476C}.

The reionization epoch of helium, by contrast, is far less well
constrained. This is primarily because of the need to go to space to
measure the intergalactic Lyman series absorption of intergalactic gas
in the spectra of high redshift QSOs. The current constraints place
the helium reionization epoch, when helium becomes nearly fully doubly
ionized, at $z\gsim3$ \citep{Reimers05}. Because of the high
photoelectic threshold energy of \HeII, it is expected that helium was
reionized by QSOs rather than stars, most likely in the redshift
interval $z=3-4$ \citep{MM94, Meiksin05, 2009ApJ...694..842M}.

Measurements of the \HeII\ \Lya\ optical depth show the optical depth
rises rapidly approaching $z\lsim3$ with an increasing amount of
patchiness \citep{Zheng04, 2006A&A...455...91F}. By $z\approx3.4$, the
optical depth is immeasurably large \citep{2008ApJ...686..195Z}. The
limits are not yet strong enough to demonstrate the epoch of helium
has been detected, but the rapid rise in optical depth and degree of
patchiness are suggestive of an approach to the helium reionization
epoch. This interpretation, however, is not unique:\ it may also
indicate a diminishing attenuation length of \HeII-ionising photons,
resulting in large fluctuations in the \HeII-ionising metagalactic
background due to large local Poisson fluctuations in the numbers of
QSOs contributing to the background \citep{2007arXiv0711.3358M}. The
uncertainty in QSO counts and the spectral shape of high redshift QSOs
still does not allow much earlier reionization, up to $z\lsim5.5$, to
be excluded \citep{Meiksin05}.

In this paper, we examine an alternative means of probing the epoch
when helium in the intergalactic medium (IGM) surrounding a QSO was
reionized:\ detections of a heightened IGM temperature above the
post-reionization equilibrium value. The temperature following full
helium reionization will reach high values as the helium ionization
front (I-front) passes, eventually establishing a lower temperature in
overdense regions as the gas achieves thermal equilibrium
\citep{Meiksin94, 2007MNRAS.380.1369T, 2009ApJ...694..842M}.  If the
IGM near the QSO is polluted with metal lines, then a direct
determination of the gas temperature may be made by combining the
measured linewidths of the metals, provided they are adequately
resolved \citep{1996ApJ...467L...5R, 2007arXiv0711.3358M}. Because of
the low abundances of intergalactic metals, however, measurements have
so far been restricted to high \HI\ column density systems. The advent
of large telescopes, like the Thirty Meter Telescope\footnote{\tt
  http://www.tmt.org} or an Extremely Large Telescope\footnote{\tt
  http://www.eso.org/sci/facilities/eelt/}, may permit temperature
measurements to be extended to much smaller column densities.

Whilst such direct temperature estimates are currently limited to high
column density absorbers, the reionization of helium by a QSO will
induce a ``thermal proximity effect'' in which the increased
temperature near the QSO will be reflected by broadened hydrogen, as
well as helium and metallic, absorption lines. The effect has been
exploited as a means of probing the epoch of full helium reionization
through the global statistical effect on the distribution of the \HI\
Doppler parameters for systems throughout the IGM. Rises in the lower
cutoff of the Doppler parameter distribution have been reported by
\citet{2000ApJ...534...41R} and \citet{Schaye00}, and interpreted,
based on calibrations with numerical simulations, as evidence for a
sharp temperature jump and reduction in the effective polytropic index
towards isothermality near $z=3.0$, due to the onset of \HeII\
reionization. A similar conclusion was reached by
\citet{2002ApJ...564..153Z} in a wavelet-based analysis of QSO
spectra. Global reionization simulations and semi-analytic estimates
allowing for distributed sources, however, are not clearly in accord
with such sharp changes in the absorption line parameters. The results
suggest only a gradual rise in the mean IGM temperature should occur
during helium reionization \citep{2009ApJ...694..842M,
  2009MNRAS.395..736B}. Searches for isolated regions of increased
temperature, as may arise around a helium-ionising QSO, have been only
partially successful. \citet{2002ApJ...564..153Z} argues at least ten
QSO sitelines are required to probe temperature fluctuations at a
statistically significant level. Using a similar method,
\citet{2002MNRAS.332..367T} find a region $10^4\kms$ across suggestive
of a statistically significant elevated temperature in a single QSO
siteline out of eight examined.

A more direct means of searching for the reionization signature is to
target the region around specific QSOs. The temperature of the IGM gas
surrounding a QSO may be probed both using lines of sight towards the
QSO and nearby on the sky towards background QSOs. Surveys of QSO
neighbours with follow-up spectroscopy along multiple lines of sight
are currently underway in an attempt to measure the transverse
hydrogen ionization proximity effect and to use transverse spatial
correlations in the hydrogen \Lya\ forest as a constraint on the
vacuum energy density of the Universe \citep{2008ApJS..175...29M}. The
detection of a thermal proximity effect would provide additional
support for unification models of Active Galactic Nuclei (AGN). In
these models, the variety of AGN activity observed, from radio
galaxies to QSOs and from Seyfert 1s to Seyfert 2s, is a consequence
of geometry:\ an obscuring dusty torus or other obstacle blocks the
optical/UV radiation from the central engine along some lines of
sight, possibly re-radiating it in the infra-red. The observed
character of the object then depends on the viewing angle. While good
evidence exists for elements of the model \citep{2005SSRv..119..355V,
  1993ARA&A..31..473A}, direct evidence for optical/UV radiation
transverse to the line-of-sight of a putative obscured QSO or Seyfert
1 is generally lacking. Prior to the completion of helium reionization
in the IGM, the broadening of \Lya\ absorption features transverse to
the line of sight toward bright radio or infra-red galaxies, Type II
QSOs, or Seyfert 2 galaxies would provide evidence for intense
photoionising UV radiation beamed transverse to the line of sight, as
predicted by the unification models.

The purpose of this paper is to use numerical reionization simulations
to quantify the impact of \HeII\ reionization by a QSO on the
temperature and absorption line widths produced in the surrounding
gas. The simulations are performed using a radiative transfer code
coupled to an $N$-body code to model the evolution of the IGM. The
method was previously described in \citet{2007MNRAS.380.1369T}. The
details of the simulation volume and sources are given in
\sect{sec:Simulations}. Results of the simulations are provided in
\sect{sec:Results} and discussed in \sect{sec:Discussion}. Our
conclusions are summarised in \sect{sec:Conclusions}. In an
Appendix, we describe modifications to the radiative tranfer code
implemented to simulate reionization by a central source.

\section{Simulations}
\label{sec:Simulations}

\begin{figure*}
\includegraphics{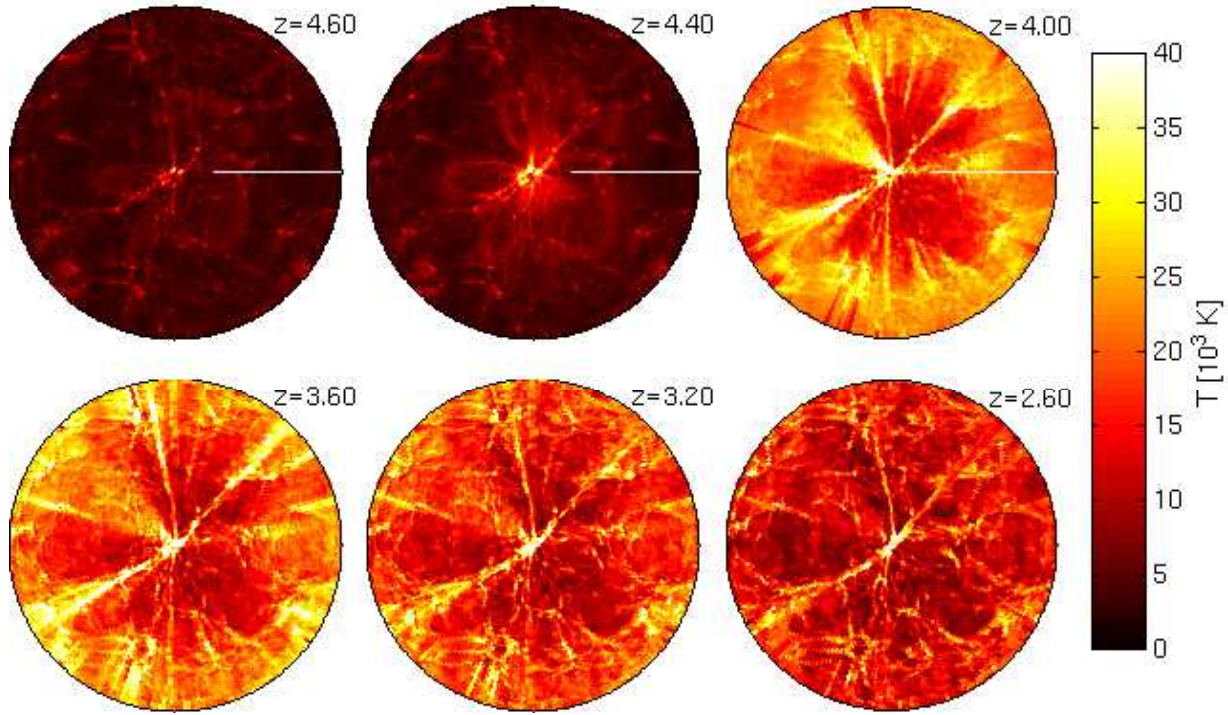}
\caption{The evolution of the gas temperature surrounding a central
  radiation source which evolves from a starburst spectrum to QSO
  (power-law) over the redshift interval $4.5 > z > 3.5$. The source
  is at a density maximum in the simulation volume. The images show
  the rapid increase in temperature following the turn on of the QSO.
  The polar grid radius is $25/2^{1/2}~h^{-1}$~Mpc (comoving).
}
\label{fig:temperature}
\end{figure*}

We use {\tt PMRT} \citep{2007MNRAS.380.1369T} to simulate the
expanding ionization front produced by a central source. The
simulation code is the merger of a Lagrangian particle-mesh (PM) code
\citep{MWP99}, and a grid-based radiative transfer (RT) code based on
a probabilistic photon-conserving algorithm
\citep{1999ApJ...523...66A, BMW04}. The method assumes the baryons
trace the dark matter. The code is thus not able to cope with sudden
increases in the pressure forces as the gas is heated and driven out
of shallow potential wells or heated by shocks in collapsing
structures \citep{1988ApJ...324..627B, Meiksin94}. Whilst
pseudo-hydrodynamical methods are no replacement for full
hydrodynamics computation for high-precision comparisons with data,
comparisons with full hydrodynamics computations show that the method
recovers the distribution of Doppler parameters to high accuracy,
although the median Doppler parameter may be slightly broadened by
$1-2\kms$ \citep{MW01}, which is an intrinsic limitation to our
estimates. An alternative method including an enthalpy-based
pseudo-pressure term similar to the HPM scheme of
\citet{1998MNRAS.296...44G} was found to broaden the lines even
more. For this reason we adhere to the more straightforward
implementation of \citet{MW01}. In future full hydrodynamics
computations including radiative transfer would be preferable, and the
authors are developing such a scheme. The pseudo-hydrodynamics scheme
none the less does have advantages over a fully hydrodynamical one:\
it runs much more quickly without the hydrodynamics overhead, and the
reduced memory requirements permit simulations in larger boxes at the
high spatial resolution required to resolve the absorption
systems. The latter is important to ensure the simulations both
capture the large scale modes required to converge on the Doppler
parameters and to resolve the internal structure of the systems
\citep{Theuns98, 1999ApJ...517...13B, MW01}.

The central QSO is modelled as a starburst that gradually develops
into a hard QSO spectrum. The starburst spectrum was produced by {\tt
  P\'EGASE}\footnote{\tt http://www2.iap.fr/users/fioc/PEGASE.html}
\citep{FR97} for a galaxy 30 Myr after a burst of Pop III (zero
metallicity) star formation. The starburst spectrum has a luminosity
at the hydrogen Lyman edge of $10^{24}\,{\rm W\,Hz^{-1}}$. It is able
to ionize hydrogen and some neutral helium to singly ionized helium,
but has negligible flux above the ionization threshold of singly
ionized helium. The spectral index of QSO sources at these high
energies, particularly above the \HeII\ Lyman edge which is most
relevant for this work, is virtually unconstrained by
observations. Using {\it HST} QSO spectra covering primarily the
redshift range $0.3<z<2.3$, \citet{2002ApJ...565..773T} report a steep
source spectral index of $\alpha_S=1.76\pm0.12$ for a fit over the
spectral range $0.8-1.8$~Ry, still a factor of two short of the \HeII\
Lyman edge. By contrast, using {\it FUSE} data for QSOs at $z<0.67$,
\citet{2004ApJ...615..135S} find a spectral index of
$\alpha_S=0.56^{+0.28}_{0.38}$ fit over the similar spectral range
$0.8-1.4$~Ry. A fair fraction of the QSOs have quite hard spectra with
$\alpha_S<0.$ The origin of the discrepancy with the {\it HST} sample
is unclear. The {\it FUSE} sample is somewhat fainter, although there
is substantial overlap in the QSO luminosities with the {\it HST}
sample. A correction for any intervening Lyman limit systems and a
statistical correction for intervening intergalactic absorption are
applied to the {\it HST} spectra, which is unnecessary for the lower
redshift {\it FUSE} sample, so that the {\it FUSE} sample may be a
truer reflection of the intrinsic QSO spectra at these
energies. Alternatively, it may be that QSOs harden at lower redshift,
although \citet{2002ApJ...565..773T} find no such effect in their
data. Even if most QSO spectra tend to be softer at higher redshifts,
for a given luminosity at the \HeII\ Lyman edge QSOs with harder
spectra will provide a greater number of ionizing photons, and so they
may still dominate the reionisation of helium in the IGM. The QSO
spectrum in our simulations is approximated as an $\alpha_S=0.5$ power
law at energies above the \HI\ Lyman edge. Although a hard spectrum is
suggested by the {\it FUSE} data, our choice is primarily motivated by
an attempt to address the discrepancy between the measured \HI\
Doppler parameters and the predictions of simulations. Detailed
comparisons between simulations and data suggest the simulations
substantially underpredict the temperature of the gas \citep{Theuns98,
  1999ApJ...517...13B, MBM01}. A harder spectrum will produce a higher
temperature. If still inadequate, then additional sources of heating
or line-broadening (such as winds) may be required. Ultimately an
assessment of the impact of \HeII\ reionization on the gas
temperatures will require improved knowledge of the intrinsic spectra
of QSOs, particularly at energies above the \HeII\ Lyman edge.

\begin{figure}
\scalebox{0.7}{\includegraphics{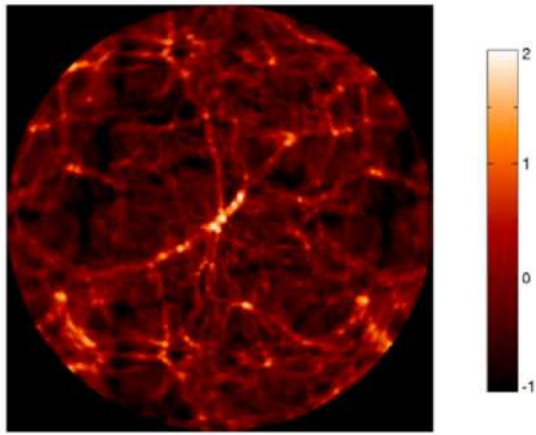}}
\caption{The density contrast
  $\delta\rho=(\rho-\langle\rho\rangle)/\langle\rho\rangle$ at $z=2.6$
  surrounding a central source positioned at a density maximum in the
  simulation volume. The shading ranges from $\delta\rho=-1$ (black)
  to $\delta\rho>100$ (white). The polar grid radius is
  $25/2^{1/2}~h^{-1}$~Mpc (comoving).
}
\label{fig:overdensity}
\end{figure}

The QSO spectrum is normalised to provide a peak hydrogen-ionising
photon production rate half that of the starburst, with a peak
luminosity at the hydrogen Lyman edge of $2.6\times10^{22}\,{\rm
  W\,Hz^{-1}}$. Three transitions from starburst spectrum to QSO were
simulated:\ a gradual transition over the redshift intervals $z=5.5$
to $z=4.5$, $z=4.5$ to $z=3.5$ and $z=3.5$ to $z=2.5$. The starburst
spectrum was turned on at $z=8$ in all cases. In principle, an
ionization front representing the accumulated ionising photons from a
population of galaxies could have been swept across the simulation
volume, however we find that the IGM temperatures have relaxed by the
time the QSO spectrum turns on, without any memory of the central
source, so that initiating hydrogen ionization by a central source
adequately produces the expected temperature structure of the IGM
prior to the turn-on redshift of the QSO spectrum. In post-processing
the spectra, we have added a uniform hydrogen ionization rate to match
recent estimates.

To approximate the expected environment of a QSO, we computed mass
overdensities in the simulation after smoothing the PM distribution at
$z=3.2$ over a comoving scale of $0.5h^{-1}$~Mpc, and placed the
source at an overdensity peak corresponding to a mass of
$1.5\times10^{13}\,{\rm M_\odot}$, consistent with recent estimates of
QSO dark matter halo masses based on QSO space density and clustering
statistics \citep{2007AJ....133.2222S, 2008MNRAS.390.1179W}. In order
to gauge the importance of the environment to the resulting properties
of the IGM, we also computed a set of models with the source placed at
the centre of an underdense region. Qualitative effects found in
common may then be expected for any source location. Since galaxies
are also found in underdense regions, the simulations may in fact
describe a plausible helium reionization scenario, since an
appreciable fraction of galaxies may have undergone an AGN phase at
some stage in their lives so that helium in some underdense regions
may have been reionized by a local source residing within them.

\begin{table}
\caption{Identifiers for the various source models used in this paper.}
\begin{tabular}{rccc}
\hline & \multicolumn{3}{c}{$z_{\rm on}$}\\ Environment & 3.5 & 4.5 & 5.5 \\
            \hline High density & HD3.5 & HD4.5 & HD5.5 \\
            \hline Low density & LD3.5 & LD4.5 & LD5.5 \\ \hline
\label{table.identifiers}
\end{tabular}
\end{table}

\begin{figure*}
\includegraphics{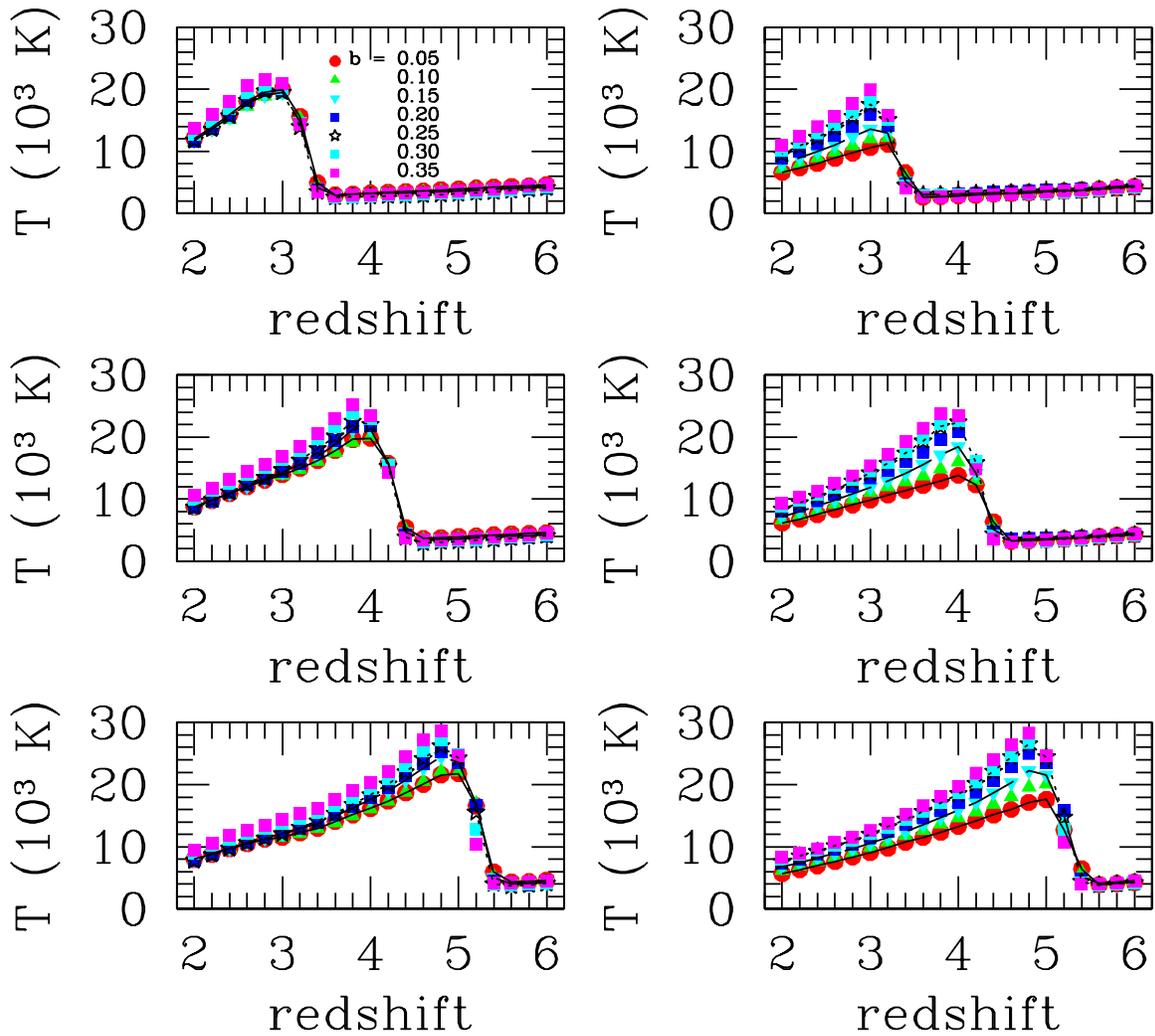}
\caption{The evolution of azimuthally-averaged line-of-sight gas
  temperatures around a central radiation source. The left column
  corresponds to a source placed at a density maximum while for the
  right column the source is placed at the centre of a void. From top
  to bottom, the source evolves from a starburst spectrum to QSO
  (power-law) over the redshift intervals $5.5 > z > 4.5$, $4.5 > z >
  3.5$ and $3.5 > z > 2.5$, respectively. The different curves
  correspond to constant comoving impact parameters for the lines of
  sight of $b=0.05$ (solid line), 0.15 (long--dashed line) and 0.25
  (short--dashed line), in units of the polar grid diameter. The
  temperatures show a monotonically increasing trend with impact
  parameter after the \HeIII\ I-fronts pass. The polar grid radius is
  $25/2^{1/2}~h^{-1}$~Mpc (comoving).
}
\label{fig:los-temperature}
\end{figure*}

\begin{figure*}
\includegraphics{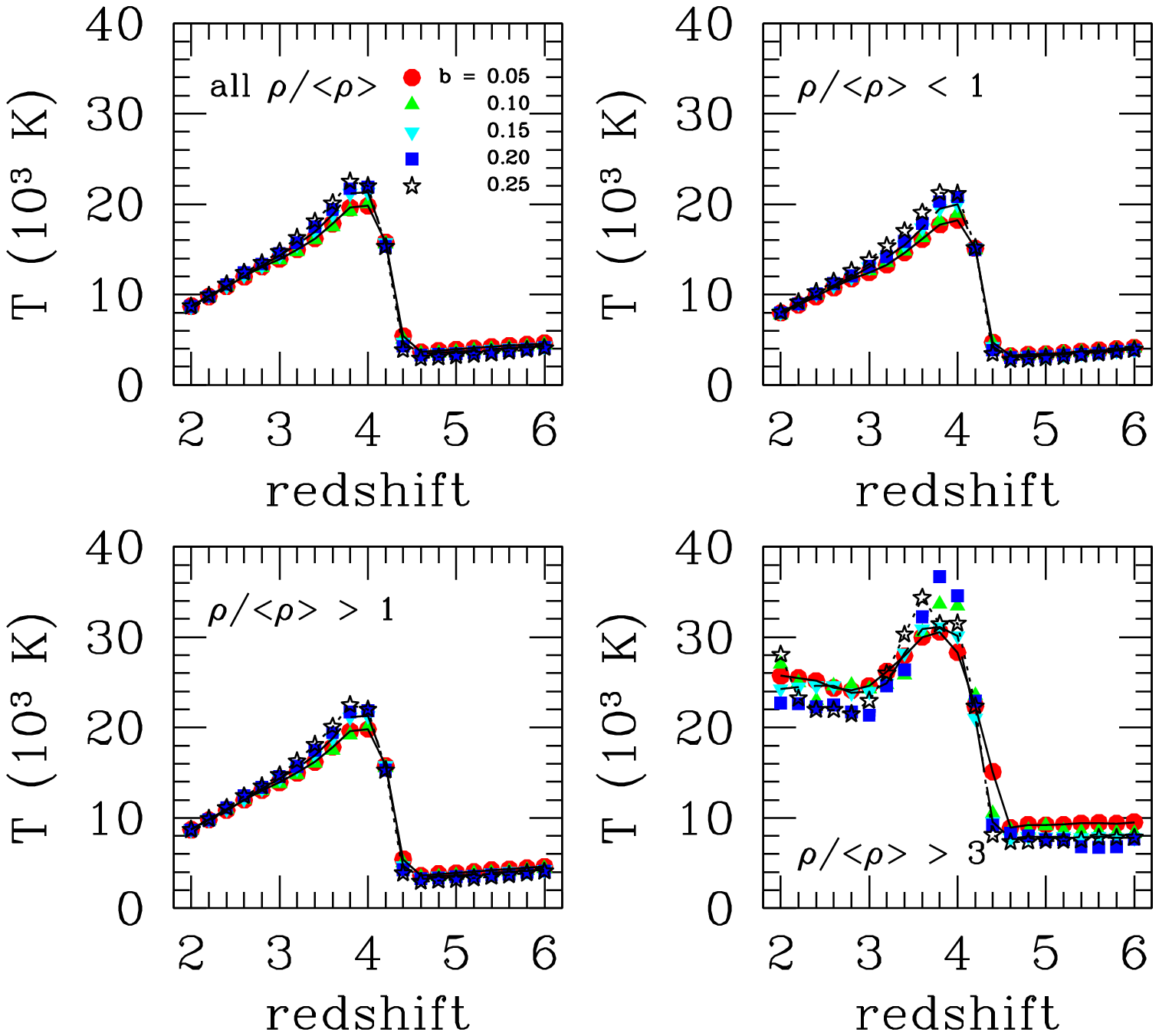}
\caption{The evolution of the azimuthally-averaged line-of-sight
  median gas temperature for different over-density cuts, for run
  HD4.5. The different curves correspond to projected separations of
  $b=0.05$ (solid line), 0.15 (long--dashed line) and 0.25
  (short--dashed line), in units of the polar grid diameter.
}
\label{fig:T-overdensities}
\end{figure*}

For brevity, the labels listed in \tab{table.identifiers} will be used
henceforth to identify the various simulations with their combination
of source location and source power-law turn-on redshift.

All simulations, unless stated otherwise, were performed in a
$(25h^{-1} \Mpc)^3$ comoving volume. A \LCDM\ model was assumed
\citep{Spergel03}, with parameters: $h=0.71$, $\Omega_b h^2 = 0.022$,
$\Omega_m = 0.268$ and $\Omega_v = 0.732$, where $\Omega_b$,
$\Omega_m$, and $\Omega_v$ have their usual meanings as the
contributions to $\Omega$ from the gas, all matter, and the vacuum
energy, respectively. We adopt a helium fraction by mass of $Y=0.235$.

The initial density perturbations were created by displacing a uniform
grid of $512^3$ particles in the box using the Zel'dovich
approximation. The initial power spectrum of the density fields was a
{\it COBE}-normalised power law with index $n=0.97$. The same initial
conditions were used for all simulations. Since there is no feedback
from the RT to the PM code, all the runs have identical gas
densities. The simulations were evolved to a redshift of 2.

The reionization computations were carried out on a superimposed polar
grid. Details of the scheme are provided in the Appendix. The finite
travel time of light is included in the propagation of the radiation
field from the source. Because of the intensive computational demands
required for accurate temperature determinations associated with
reionization, the regions ionized are restricted to slabs $500
h^{-1}$~kpc (comoving) thick and $25/ 2^{1/2} h^{-1}$~Mpc (comoving)
in radius. This produces a small amount of periodic overlap in the
density field at the edges of the polar grid along its cardinal
directions, necessary to maximise the usage of the cubic simulation
volume.

\section{Simulation results}
\label{sec:Results}

\begin{figure}
\scalebox{0.6}{\includegraphics{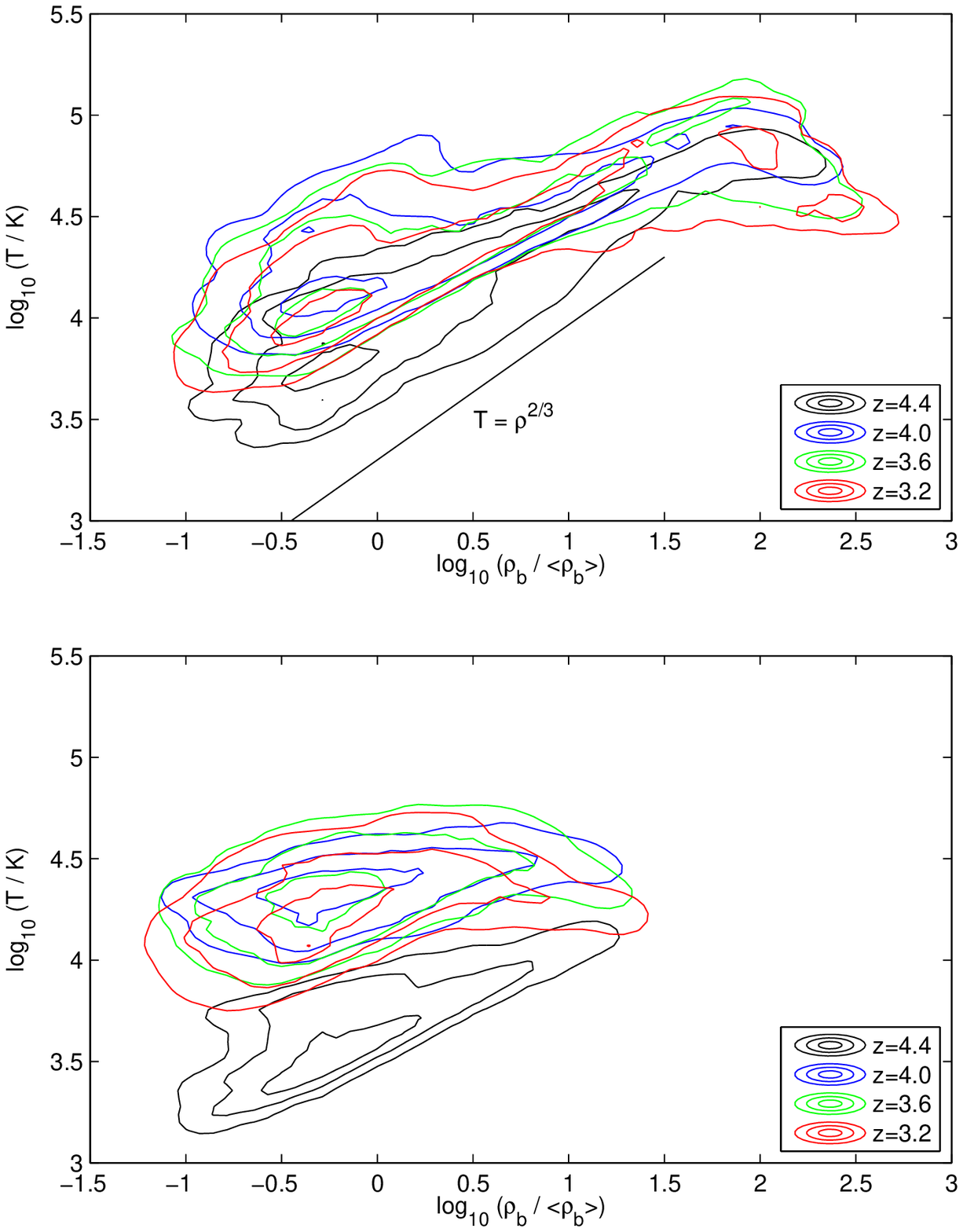}}
\caption{The evolution of the relation between gas density and
  temperature in the vicinity of a source placed at a density
  peak. The dependence is split between the inner half radius of the
  polar grid (top panel) and outer half radius (bottom panel). The
  contour levels correspond to probability density contours
  per logarithmic intervals in temperature and overdensity of
  $10^{-0.5}$, $10^{0.5}$ and $10^{1.5}$.
}
\label{fig:overdens_vs_T}
\end{figure}

\begin{figure}
\scalebox{0.6}{\includegraphics{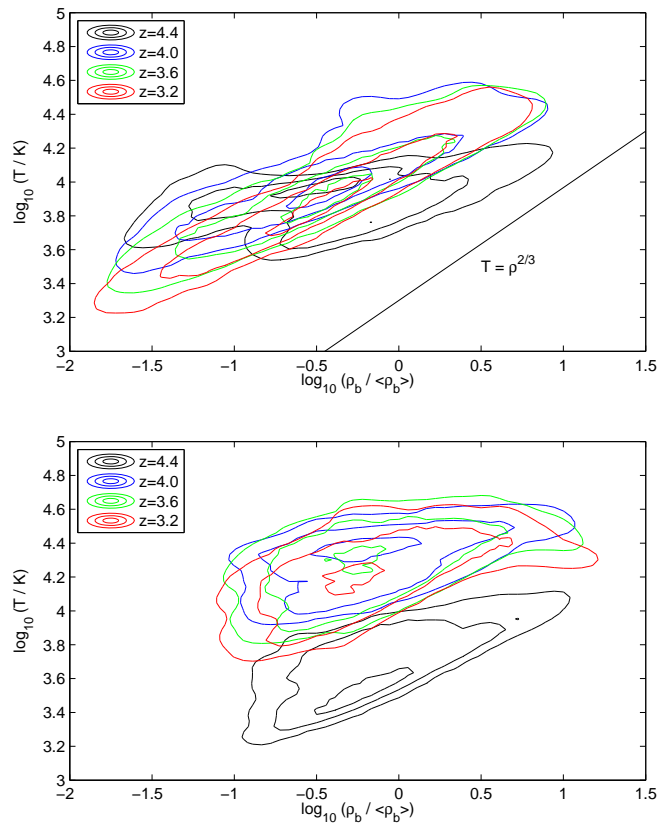}}
\caption{The evolution of the relation between gas density and
  temperature in the vicinity of a source placed at a density
  minimum. The dependence is split between the inner half radius of
  the polar grid (top panel) and outer half radius (bottom panel). The
  contour levels correspond to probability density contours
  per logarithmic intervals in temperature and overdensity of
  $10^{-0.5}$, $10^{0.5}$ and $10^{1.5}$.
}
\label{fig:overdens_vs_T-min_centred}
\end{figure}

\subsection{Physical properties of the IGM}
\label{sec:IGMphysics}

We focus on the thermal state of the gas surrounding the central
source before and after the \HeIII\ I-front passes. The evolution of
the temperature in a slice through the simulation is shown in Figure
\ref{fig:temperature}. The role played by dense structures surrounding
the source in restraining the growth of the \HII\ and \HeIII\ I-fronts
is apparent in the highly anisotropic temperature patterns. For
comparison, the density field at $z=2.6$ is shown in
Figure~\ref{fig:overdensity}. (Since the overdensities change very
little over the redshift interval $2.6<z<4.6$, only the image at
$z=2.6$ is shown.) The gas temperature increases sharply behind the
\HeIII\ I-fronts, then relaxes to lower values as the gas achieves
equilibrium between photoionization heating and atomic radiative
losses. As a consequence, the highest temperatures tend to lie
furthest from the source in regions behind the \HeIII\ I-front. Thus,
whilst the gas temperature is increased within the \HeIII\ region
surrounding the QSO compared with the gas temperature outside, the
temperature of the gas within the \HeIII\ region decreases towards the
QSO itself. As will be discussed in \sect{sec:Discussion} below,
another important contributing factor to the trend is the hardening of
the radiation field by the intervening gas between the source and the
more distant regions.

Figure~\ref{fig:los-temperature} shows azimuthal averages of the
median gas temperature along lines of sight drawn around the source
with varying impact parameter. Once the power-law source turns on, a
steep rise in the temperature with decreasing redshift is found. The
rise occurs over a much narrower redshift interval than the transition
redshift interval of the source spectrum as the source becomes
sufficiently hard to photoionize \HeII. The associated photoionization
heating occurs over a redshift interval of approximately $\Delta
z\simeq0.4$, with the temperature peaking at the centre of the
transition redshift interval. The median averages show a trend of
increasing temperature with increasing impact parameter of the lines
of sight after helium is fully ionized. The median temperatures
approach convergence with time as photoionization heating comes into
equilibrium with atomic radiation losses and adiabatic expansion
cooling.

Similar trends are found for both underdense and overdense regions, as
shown in Figure~\ref{fig:T-overdensities}, with peak temperatures of
$T\approx20\times10^3$~K after \HeII\ reionization, diminishing to
$T\approx10^4$~K by $z=2$. (Since the projected radii $b>0.25$ begin
to sample the rising density field in the next periodic image of the
simulation volume, subsequent plots are truncated to $b\le0.25$.) At
high overdensities, however, the trend is less clear and appears to
reverse after helium reionization as a function of impact parameter,
as shown by the panel for $\rho/\langle\rho\rangle>3$. This density
threshold corresponds to systems with \HI\ \Lya\ optical depths of
about unity, approximately the minimum for which metal absorption
systems have been detected (eg, \citet{2008ApJ...689..851A}). For this
density cut, after the initial rise the temperature settles to roughly
constant values depending weakly on the impact parameter of the line
of sight, but now with lower temperatures for the larger impact
parameters. The median temperatures are also found to be considerably
higher than for most of the gas. The peak median temperature following
full helium reionization reaches values of
$T\approx30-40\times10^3$~K, decreasing to $20-30\times10^3$~K as
thermal equilibrium between photoionization heating and atomic
radiative losses is established.

The trend of temperature with impact parameter is thus not a
consequence of a narrow equation of state relating the gas temperature
to the gas density. The post-ionization gas in fact shows a wide range
of temperatures at a given overdensity in reionization computations
including radiative transfer \citep{BMW04, 2007MNRAS.380.1369T,
  2009MNRAS.395..736B}. The dependence between gas temperature and
density is shown in Figure~\ref{fig:overdens_vs_T}. The relation has
been split into two regions:\ an inner region within a half radius of
the polar grid (top panel), and the complementary outer region (bottom
panel). At $z=4.4$, just after the power-law source turns on, the
temperatures begin to elevate in the inner region as \HeII\ is
photoionized. The gas temperature and density obeys an approximate
polytropic relation. This arises in underdense regions from a balance
between photoionization heating and adiabatic expansion cooling
\citep{1997MNRAS.292...27H}. At densities above $n_{\rm
  H}\approx10^{-4}\,{\rm cm^{-3}}$ (corresponding to
$\rho/\langle\rho\rangle>3-10$ for $3<z<4.5$), however, the gas
density is sufficiently high that there is time for thermal balance to
be established between photoionization heating and atomic radiative
cooling, which results in a trend of decreasing temperature with
increasing density \citep{Meiksin94}. This turnover is present in the
outer region (lower panel). The extension of the polytropic relation
to higher overdensities in the inner region reflects the role of
adiabatic compression heating within collapsing structures, with
$T\sim\rho^{2/3}$ and reaching peak temperatures of mainly
$3-5\times10^4$~K, as expected for collapsing structures on scales of
100~kpc (proper) and smaller \citep{Meiksin94}. The
temperature-density slope is slightly shallower than the adiabatic
relation (Figure~\ref{fig:overdens_vs_T}), as radiative cooling limits
the temperatures in overdense structures. At virial densities
($\rho/\langle\rho\rangle\gsim200$), for which the gas comes to rest,
atomic processes lower the temperature, creating a downward trend in
the relationship between temperature and density (upper panel of
Figure~\ref{fig:overdens_vs_T}). The highest temperature regions
($T>40000$~K) in Figure~\ref{fig:temperature} align with the densest
structures in Figure~\ref{fig:overdensity}, with the highest
temperatures in the dense structure at the centre of the grid.

By $z=4.0$, the \HeIII\ I-front has reached the edge of the simulation
volume. The temperature-density relation has relaxed to a new
polytropic relation in the inner region as adiabatic expansion cooling
lowers the temperature of the underdense gas with time (upper
panel). The turnover at high densities is consistent with the
dominance of atomic processes in establishing the temperature-density
relation, as discussed above. By comparison, in the outer region,
where the \HeII\ was more recently photoionized, the
temperature-density relation shows a pronounced flattening, gradually
settling to a polytropic relation with a slightly steeper gradient
with decreasing redshift as adiabatic cooling lowers the temperature
of the underdense gas. In both regions, a wide spread in the
relationship is found. The high median temperatures ($T>40000$~K), in
the inner region at high densities are not reached in the outer
regions because of the absence of massive halos comparable to that
situated at the centre of the slice.

Similar evolution is found for a source turning on in an underdense
environment, as shown in
Figure~\ref{fig:overdens_vs_T-min_centred}. In this case, the inner
region heats up more quickly after the source turns on, since the
\HeIII\ ionization front is able to propagate more quickly. The faster
expansion speed in the void surrounding the source results in an
increased amount of adiabatic cooling, bringing the gas to lower
temperatures in the vicinity of the source (top panel). Since the gas
does not reach high overdensities in the inner region, the high
temperatures associated with collapsing halos found for the source
placed at a density peak do not occur. Further from the source (bottom
panel), the temperature evolves similarly to the case with the source
placed in a dense environment.

\subsection{Absorption line properties of the IGM}
\label{sec:IGMabs}

Whilst isolated features in the \HeII\ \Lya\ forest have been detected
up to redshifts $z<3$ \citep{2001Sci...293.1112K}, it becomes
increasingly difficult to detect individual \HeII\ \Lya\ absorption
features at $z>3$, as observations from space are required and the
sources tend to be too faint at these wavelengths to readily measure
the spectra because of the high \HeII\ \Lya\ optical depths at these
redshifts \citep{2007arXiv0711.3358M}. We explore the possibility of
detecting the effects of \HeII\ reionization on the \HI\ \Lya\ forest
instead. The hydrogen ionization fractions were recomputed allowing
for a background metagalactic hydrogen ionization rate in addition to
that of the QSO of, in units of $10^{-12}\,{\rm s^{-1}}$ per neutral
hydrogen atom,
\begin{equation}
\Gamma_{\rm HI, -12}=
\cases{
0.8\left(\frac{4}{1+z}\right)^{1.5} &; $2<z<5.5$\cr
0.386\left(\frac{6.5}{1+z}\right)^{14} &; $5.5<z<6$,\cr
}
\label{eq:GHI12}
\end{equation}
in good agreement with recent estimates \citep{MW04,
  2004ApJ...617....1T, 2005MNRAS.357.1178B, 2008ApJ...688...85F}.
Synthetic spectra were generated by casting lines of sight to
fictitious background sources azimuthally distributed about the
central ionization source. The spectra were computed from the modified
\HI\ fractions using Eq.~\ref{eq:GHI12}, and the gas density,
temperature and peculiar velocity from the simulations, following the
procedure in \cite{MW01}. Spectral regions were examined within
$\pm1000\kms$ of the closest point to the QSO along a neigbouring line
of sight, a range sufficiently broad to produce a large number of
absorption features whilst still within the region of influence of the
central QSO. No corrections for the finite travel time of light were
made in constructing the spectra except for the time delay between the
source at any given position within the grid, which is
small. Corrections are important for an ionization front travelling at
close to the speed of light, as it will near the source when the
source first turns on. In principle, any comparison with observations
could correct for the resulting time-delay effects in the observed
spectra. The effects, however, are small over the scales presented. By
$\Delta z=0.1$ less than the turn-on redshift of the QSO, the \HeIII\
I-front is expanding at less than 0.1c. The QSO ionizes most of the
polar grid within $\Delta z=0.3$ of the turn-on redshift, with only
highly subluminally expanding patches of \HeIII\ remaining in dense
regions or regions obscured from the source by dense intervening lumps
of gas. The spectra presented are generally drawn from the simulation
either before or after the \HeIII\ I-front passes, except for a brief
interval of width $\Delta z\approx0.3$ during which the I-front is
moving sufficiently subluminally that finite speed of light
corrections are at the 10 per cent. level or smaller, and so are not
included.

\begin{figure*}
\includegraphics{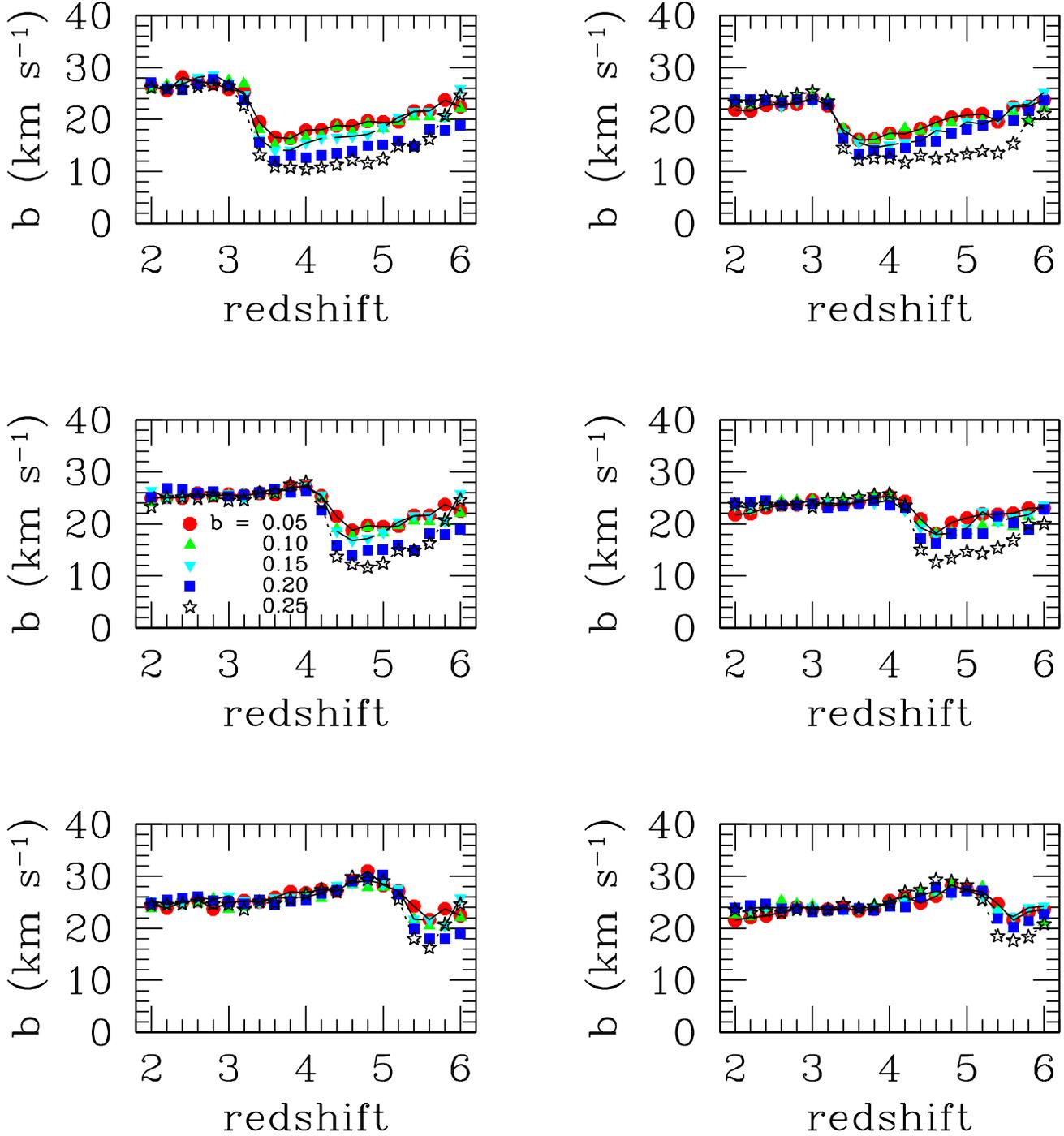}
\caption{The evolution of the azimuthally-averaged Doppler parameters
  from absorption line analyses of synthesized spectra. The different
  curves correspond to projected separations of $b=0.05$ (solid line),
  0.15 (long--dashed line) and 0.25 (short--dashed line), in units of
  the polar grid diameter.
}
\label{fig:medianb}
\end{figure*}

\begin{figure*}
\includegraphics{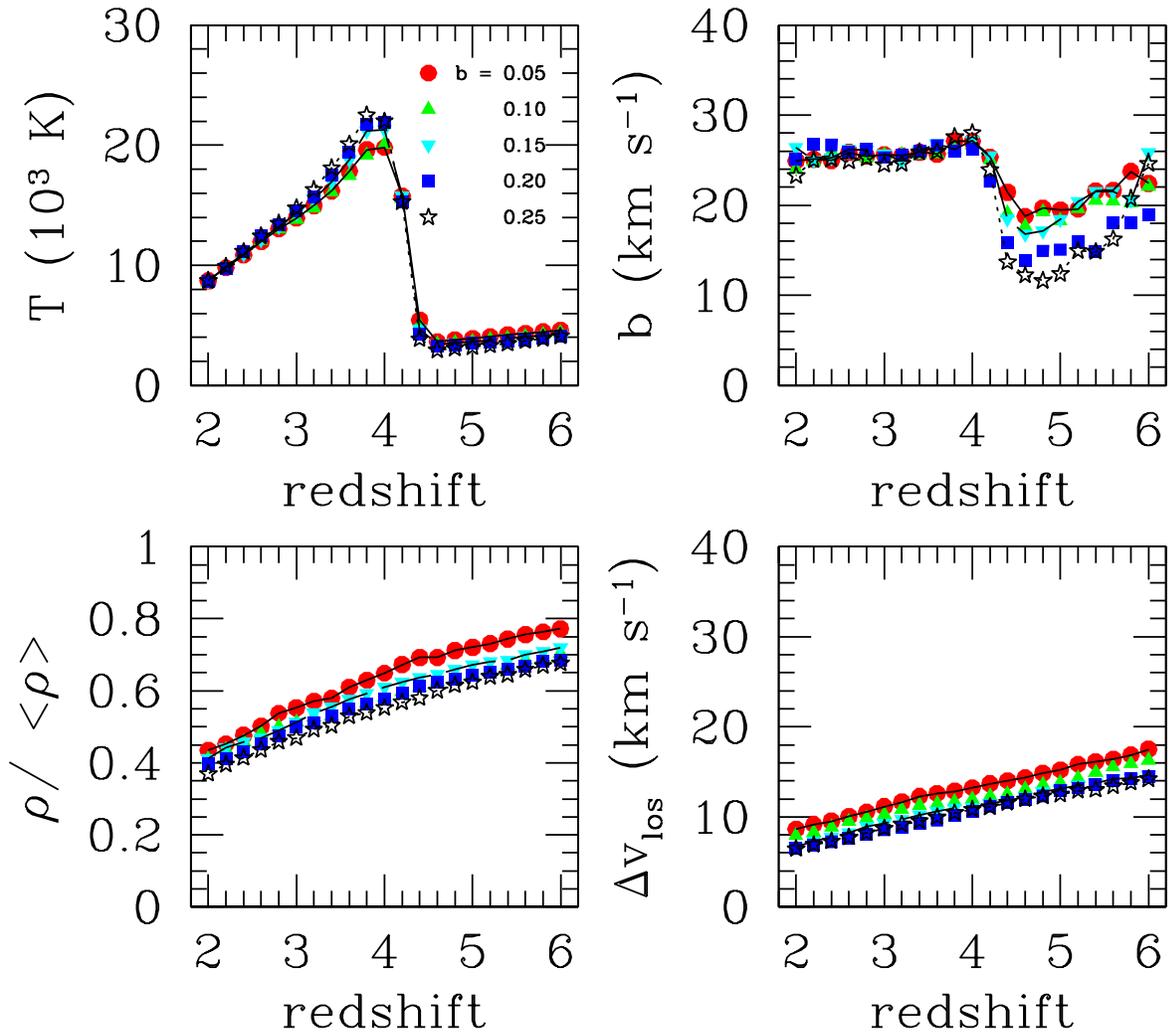}
\caption{The evolution of the azimuthally-averaged line-of-sight
  median gas temperature, over-density, peculiar velocity difference
  (over 100~kpc separations), and Doppler parameters from absorption
  line analyses to synthesized spectra, for run HD4.5. The different
  curves correspond to projected separations of $b=0.05$ (solid line),
  0.15 (long--dashed line) and 0.25 (short--dashed line), in units of
  the polar grid diameter.
}
\label{fig:T-b-rho-dvlos}
\end{figure*}

\begin{figure*}
\includegraphics{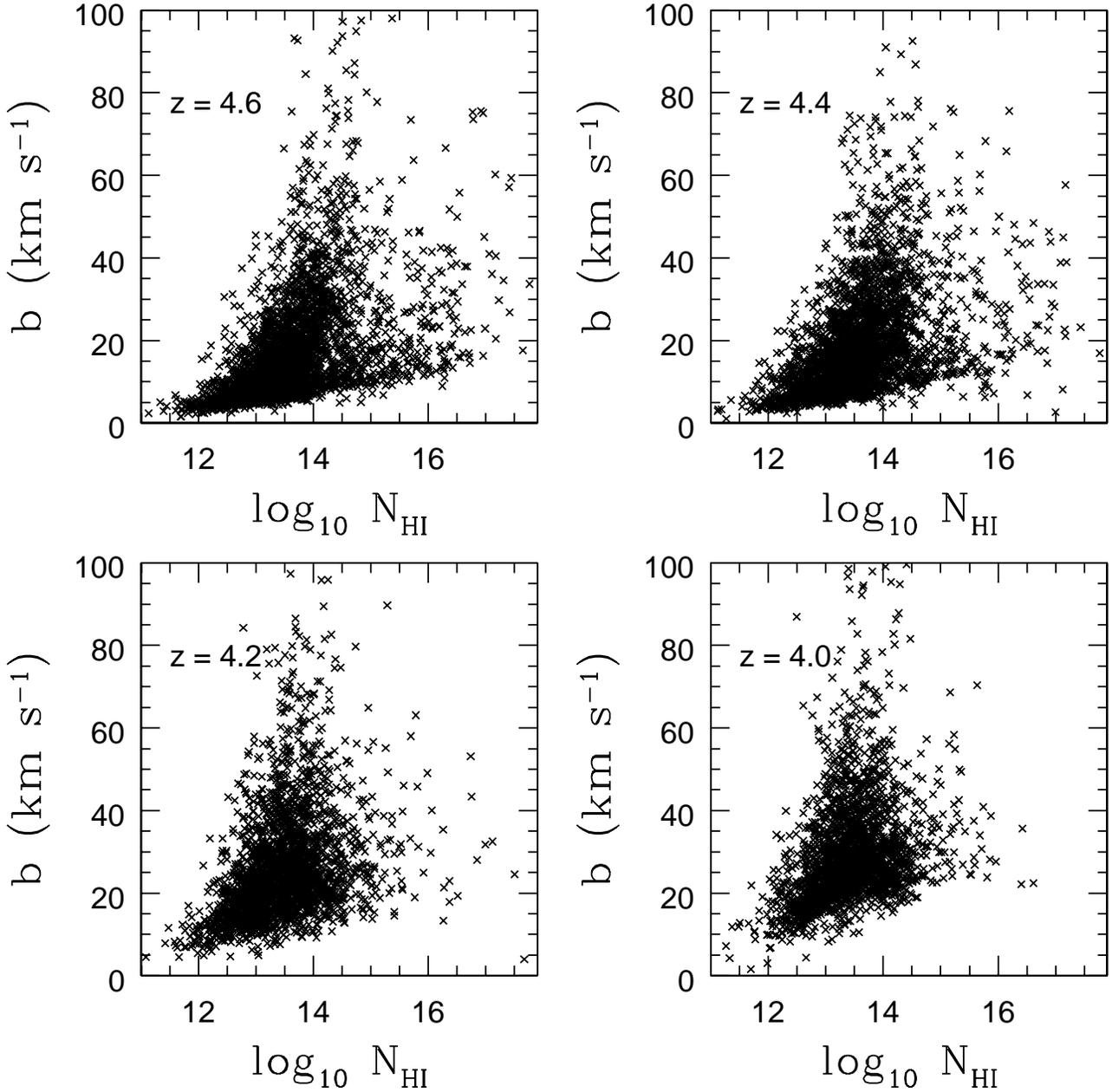}
\caption{The evolution of the \HI\ column density (in units of ${\rm
    cm^{-2}}$), and Doppler parameter joint distribution for mock
  spectra drawn from model HD4.5.
}
\label{fig:NHI-b}
\end{figure*}

\begin{figure*}
\includegraphics{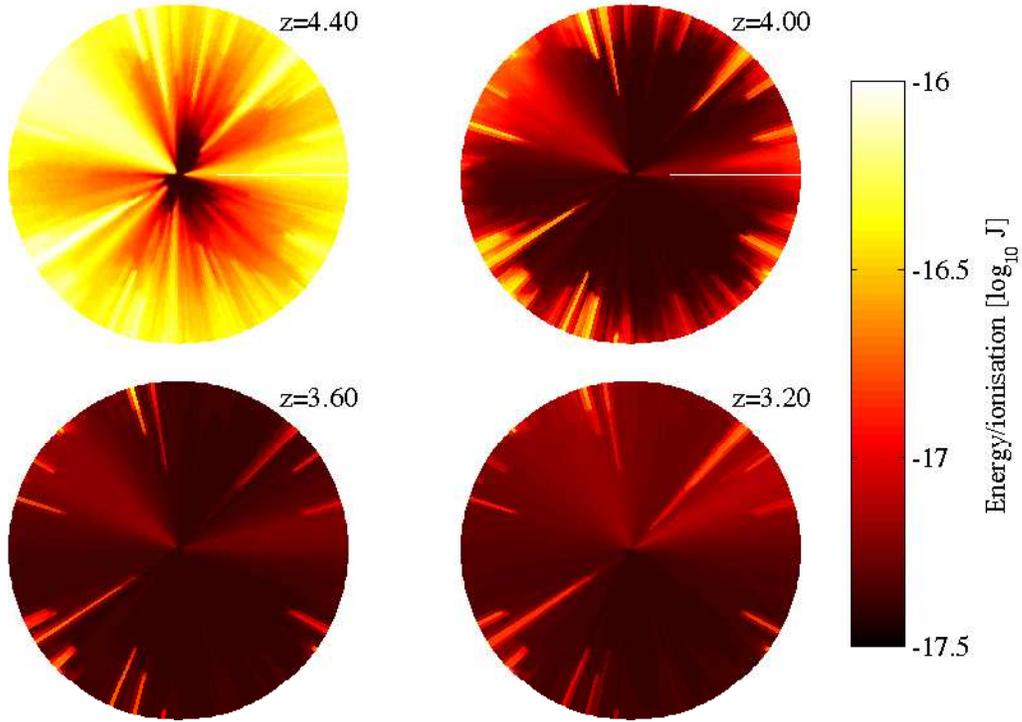}
\caption{The evolution of the rate of heat input per \HeII\ ionization
  surrounding a central radiation source which evolves from a
  starburst spectrum to QSO (power-law) over the redshift interval
  $4.5 > z > 3.5$. The source is at a density maximum in the
  simulation volume. The evolution in the rates reflects the changing
  hardness of the incident ionising radiation field as filtered by the
  IGM.
}
\label{fig:energy-per-ionization}
\end{figure*}

\begin{figure*}
\includegraphics{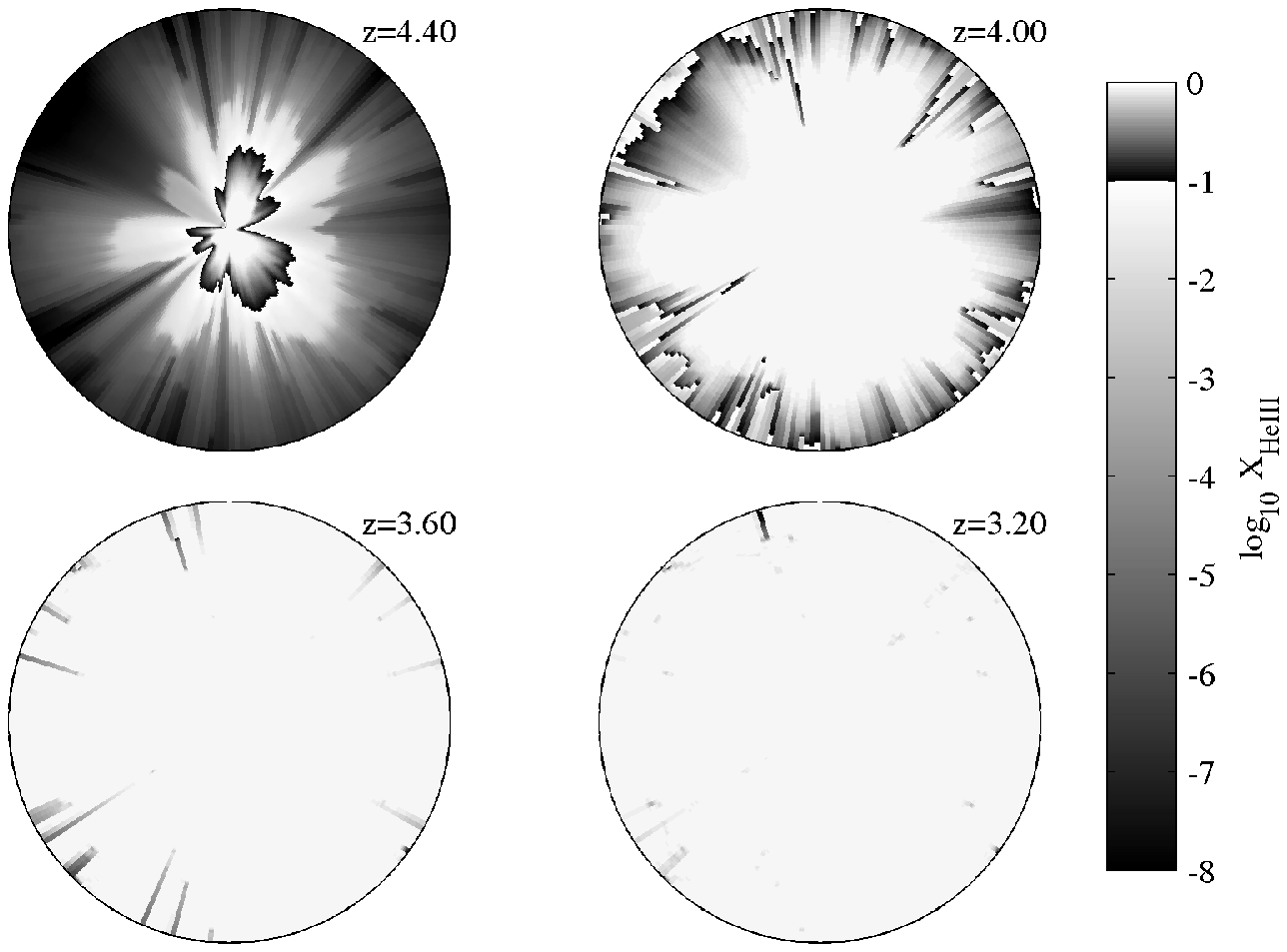}
\caption{The evolution of the \HeIII\ ionization fraction surrounding
  a central radiation source which evolves from a starburst spectrum
  to QSO (power-law) over the redshift interval $4.5 > z > 3.5$. The
  source is at a density maximum in the simulation volume.  The images
  show the rapid increase in \HeIII\ ionization once the QSO turns
  on. A double greyscale is used to capture the dynamic range of the
  \HeIII\ ionization fraction, one ranging from 0.1 (black) to 1
  (white) in the innermost regions near the source, and the other
  ranging from $10^{-8}$ (black) to 0.1 (white) in the outer regions
  in the panels at $z=4.4$ and 4.0.
}
\label{fig:HeIII-ionization}
\end{figure*}

A Voigt-profile absorption line analysis was performed on the spectra
using {\tt AUTOVP} \citep{1997ApJ...477...21D}, modified with
additional error controls to render the code more robust. Azimuthal
averages of the median values of the resulting Doppler parameters are
shown in Figure~\ref{fig:medianb}. The Doppler parameters increase as
the hard photoionization source turns on, reflecting the behaviour of
the median gas temperature in Figure~\ref{fig:los-temperature},
although with a somewhat gentler rise.

There is, however, a remarkable difference in the trend with projected
separation of the line of sight. Prior to the onset of the hard
photoionization spectrum and the subsequent full reionization of
helium, the gas temperature shows little variation with impact
parameter from the central source. The small trend indicated in the
figure is consistent with the trend with density as a function of
offset, as shown in Figure~\ref{fig:T-b-rho-dvlos}. After helium
becomes fully ionized, the temperature increases systematically with
distance from the source. By contrast, the Doppler parameter shows a
strong trend of decreasing value with increasing projected separation
from the central source prior to the onset of the hard photoionization
spectrum. The trend, along with a gradual decrease with decreasing
redshift, is apparent in the high density runs, particularly HD3.5 and
HD4.5, as well as in all the low density runs, as shown in
Figure~\ref{fig:medianb}.

After helium is fully reionized, the trends with impact parameter and
redshift gradually disappear. In fact, regardless of the onset of
helium reionization in the models, by $z=3$ the median Doppler
parameters all take on nearly the same value of about $25\kms$ in the
high density runs, in good agreement with observations
\citep{MBM01}. In the low density runs, a slightly lower median value
of about $23\kms$ is found.

The near constancy of the median Doppler parameter with both redshift
and impact parameter following helium reionization is a surprising
result. Since the Doppler parameter increases with gas temperature,
trends similar to those found for the temperature may have been
expected. The Doppler parameter, however, reflects the underlying
peculiar motions of the gas as well as the temperature, although the
relation is a complex one. The contribution of peculiar motions to the
broadening of the absorption lines is not straightforward to isolate
since the motions not only broaden the lines but displace them as
well. An attempt to construct spectra without including the peculiar
velocity field resulted in qualitatively different absorption
structures with relative displacements of line centres and
substantially different blending of lines. Since absorption line
fitting is a non-linear process, very different absorption lines
result with widths that reflect differences in the deblending of the
absorption features in addition to the absence of peculiar
motions. The total amount of absorption changes as well, making it
unclear how to renormalise the spectra for a fair comparison with the
correctly computed spectra including the peculiar velocities.

Some insight into the role of peculiar motions in broadening the
absorption lines, however, may be made by examining the structure of
the peculiar velocity field. The line-of-sight peculiar velocity
difference across a 100~kpc (proper) separation, corresponding to the
typical thickness of the absorption systems (eg, \cite{ZMAN98}), is
found to decrease with decreasing redshift, as well as to decrease the
more distant from the source, as shown for case HD4.5 in
Figure~\ref{fig:T-b-rho-dvlos}. Prior to helium reionization, when the
gas temperatures are relatively low, the Doppler parameters are
comparable to the peculiar motion differences and follow the same
trend of decreasing value with increasing impact parameter. Once
helium is fully ionized, the Doppler parameters become nearly constant
both with redshift and with impact parameter. This suggests they are
no longer probing the mean evolution in temperature and peculiar
motion of the gas, as weighted by volume, but rather the temperature
in denser regions which shows little dependence on either redshift or
distance from the source, as shown in Figure~\ref{fig:T-overdensities}
for regions with $\rho/\langle\rho\rangle>3$. It is these overdense
structures which correspond to systems with \Lya\ optical depths near
unity and above, which dominate the absorption in the spectra at
$z<4$, in contrast to the lower overdensity structures that dominate
the absorption at higher redshifts because of their higher absolute
densities \citep{ZMAN98}.

Similar trends are found for a source placed in the centre of an
underdense region. Prior to full helium reionization, the Doppler
parameter for absorption features along neighbouring lines of sight
increases with decreasing impact parameter of the line of sight
(Figure~\ref{fig:medianb}). This is found to correlate well with an
increased difference in the line of sight peculiar velocity over the
scale of 100~kpc for smaller impact parameters. At such small impact
parameters, nearly the full peculiar expansion velocity of the
underdense region will be probed, while at larger impact parameters
the contribution of the expansion velocity to the line broadening will
be reduced by the decrease in the projection along the line of
sight. After reionization, the Doppler parameters are again nearly
independent of redshift and impact parameter, suggesting again that
they are probing overdense structures for which the gas temperature
has little dependence on redshift or distance from the source.

The evolution in the \HI\ column density and Doppler parameter joint
distribution is shown in Figure~\ref{fig:NHI-b} for model HD4.5 over
the redshift range $4.6>z>4.0$. The onset of \HeII\ reionization has
two effects. It increases the median Doppler parameter, while
narrowing the width of the \HI\ column density distribution as the
recombination rate is reduced due to the increase in the gas
temperature. This enhances the reduction in the number density of high
column density systems along a line of sight due to the decrease in
the absolute density of the gas as a result of cosmological expansion
\citep{ZMAN98}. A clear envelope of minimum Doppler parameters,
increasing with $N_{\rm HI}$, is produced both before and after full
helium reionization. For systems with $N_{\rm HI}>10^{14}\,{\rm
  cm^{-2}}$, the envelope is found to rise from $b_{\rm
  min}\approx7\kms$ to $b_{\rm min}\approx10-15\kms$ following \HeII\
reionization, consistent with the increase in gas temperatures.

\section{Discussion}
\label{sec:Discussion}

Much of the temperature structure of the IGM after helium reionization
may be accounted for by the hardening of the radiation field as
filtered through the IGM during \HeII\ reionization. Approximating the
spectrum of the radiation field locally as a power law $J_{\nu}=J_{\rm
  HeII}(\nu/\nu_{\rm HeII})^{-\alpha}$, where $J_{\rm HeII}$ is the
angle-averaged intensity at the \HeII\ photoelectric threshold
frequency $\nu_{\rm HeII}$, the ionization and heating rates per
\HeII\ ion are, respectively,
\begin{eqnarray}
\Gamma_{\HeII}&=&J_{\rm HeII}\sigma_{\rm HeII}\int_{\nu_{\rm HeII}}^{\infty}
d\nu\, \frac{(\nu/\nu_{\rm HeII})^{-\alpha-3}}{h_{\rm
    P}\nu}\nonumber\\ &=&\frac{J_{\rm HeII}\sigma_{\rm HeII}}{h_{\rm
    P}(3+\alpha)},
\label{eq:GammaHeII}
\end{eqnarray}
and
\begin{eqnarray}
  G_{\HeII}&=&J_{\rm HeII}\sigma_{\rm HeII}\int_{\nu_{\rm HeII}}^{\infty} d\nu\,
  \left(\frac{\nu}{\nu_{\rm HeII}}\right)^{-\alpha-3}
  \frac{\nu-\nu_{\rm HeII}}{\nu}\nonumber\\
  &=&J_{\rm HeII}\sigma_{\rm HeII}\nu_{\rm HeII}[(2+\alpha)(3+\alpha)]^{-1},
\label{eq:GHeII}
\end{eqnarray}
where the photoelectric cross-section for ionising \HeII\ is
approximated as $\sigma_{\rm HeII}(\nu/\nu_{\rm HeII})^{-3}$, and
$h_{\rm P}$ is the Planck constant. This then gives for the expected
energy injected into the gas per ionization
\begin{equation}
  \frac{G_{\HeII}}{\Gamma_{\HeII}}=\frac{h_P\nu_{\rm HeII}}{2+\alpha}\simeq
  4.36\times10^{-18}(1+\alpha/2)^{-1}\,{\rm J}.
\label{eq:E-per-ion}
\end{equation}
Deviations from this will occur due to spectral structure within the
ionization front, especially as the mean free path of high energy
photons is long compared with that of photons just above the
photoelectric threshold, resulting in a hardening of the spectrum
through the I-front. The relation serves to provide an approximate
estimate for the expected amount of energy injection and the
sensitivity to the hardness of the effective spectral index within the
ionization front, which may even achieve negative values ($\alpha<0$)
in regions for which more high energy photons are able to penetrate
than low.

The evolution of the energy per ionization is shown in
Figure~\ref{fig:energy-per-ionization}. At early times after the
source just turns on (the panels for $z=4.4$ and $z=4.0$), heating
rates per ionization more than an order of magnitude greater than
Eq.~(\ref{eq:E-per-ion}) for $\alpha=0.5$ (the source spectral index)
are found. The helium in these regions is still largely in the form of
\HeI, with the \HeIII\ region just beginning to emerge from the source
at $z=4.4$, and not quite reaching the edge of the polar grid by
$z=4.00$, as shown in Figure~\ref{fig:HeIII-ionization}.

\begin{figure}
\scalebox{0.5}{\includegraphics{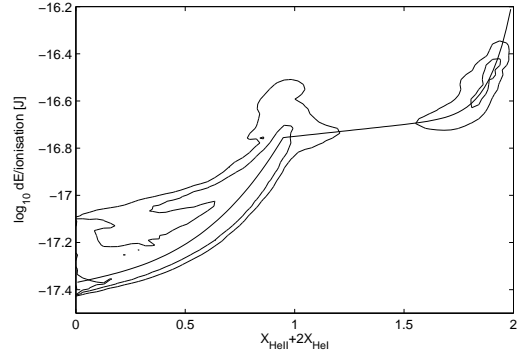}}

\caption{The energy per \HeII\ ionization as a function of the \HeI\
  and \HeII\ ionization fractions, $x_{\rm HeI}$ and $x_{\rm HeII}$,
  respectively, forming narrowly-defined contours. The dependence is
  well-parametrised as a function of $x_{\rm HeII} + 2x_{\rm HeI}$
  (thin line). The contour levels are spaced by a decade. The left
  contours correspond to $x_{\rm HeI}=0$. The data are for $z=4.0$.
}
\label{fig:E_per_ion_parametrised_contour}
\end{figure}

Whilst the \HeII\ ionization fraction $x_{\rm HeII}$ evolves rapidly
within the ionization front, it is found that the heating rate per
ionization is nearly constant for a given value of $x_{\rm HeII}$. This
is illustrated in Figure~\ref{fig:E_per_ion_parametrised_contour}.
The energy per ionization is well-described by
\begin{eqnarray}
  \frac{G_{\rm HeII}}{\Gamma_{\rm HeII}} &=& 10^{-17.37 + t/5 + t^3
    /2}~{\rm J}\qquad\quad ;t \le 0.9475\nonumber\\
  &=& 10^{-16.85 + t/10 + (t-1)^{10} /2}~{\rm J}\quad ;t >  0.9475
\label{eq:dEdI}
\end{eqnarray}
where $t = x_{\rm HeII} + 2 x_{\rm HeI}$. The right region corresponds
to gas that is ionising from \HeI\ to \HeII, whilst the left region
shows the level of \HeII\ as the helium becomes fully ionized. The
bridge between the regions indicates the rapid depletion of \HeI\ as
it is converted into \HeII. As the \HeII\ is ionized into \HeIII, the
energy per ionization rapidly declines as the effective ionising
spectrum softens towards the intrinsic spectral shape of the source.

The expected boost in the gas temperature, allowing for the injected
energy to be shared by all the species present, is
\begin{eqnarray}
  \Delta T &=& \frac{2}{3k_{\rm B}}\frac{1}{2.23}\frac{n_{\rm HeII}}{n_{\rm
      H}}h_{\rm P}\nu_{\rm HeII}\nonumber\\
  &\simeq&16800\,x_{\rm HeII}\left(\frac{G_{\HeII}/\Gamma_{\HeII}}{10^{-17}\,{\rm
        J}}\right)\,{\rm K},
\label{eq:DeltaT}
\end{eqnarray}
where $k_{\rm B}$ is the Boltzmann constant. Comparison with
Figure~\ref{fig:energy-per-ionization} shows that the rises in the
median temperature in Figure~\ref{fig:los-temperature} following full
helium ionization are of the magnitudes expected, although the actual
local increase may be much larger, depending on the evolution of the
\HeII\ ionization fraction $x_{\rm HeII}$ within the ionization
front. The gas will also rapidly cool in dense regions due to
radiative losses, and in underdense regions due to adiabatic expansion
cooling. By $z=4.00$, the highest heating rates are confined to the
periphery of the polar grid (Figure~\ref{fig:energy-per-ionization}),
where pockets of $x_{\HeII}$ continue to survive in shadows cast by
overdense clumps blocking the source, as shown by a comparison between
Figures~\ref{fig:HeIII-ionization} and \ref{fig:overdensity}. In the
inner regions, where the helium is fully ionized, the gas is able to
cool through radiative losses and the temperature lowers, leaving the
peripheral regions at relatively higher temperatures. This is the
origin of the trend of increasing temperature with impact parameter
shown in Figure~\ref{fig:los-temperature}. High heating rates are
found to persist down to $z=3.60$ even after most of the helium is
fully ionized, and indeed increase by $z=3.20$ as dense clumps
continue to gather near the centrally placed source further filtering
the radiation field, demonstrating that radiative transfer effects may
continue to affect the temperature of the gas in distant regions even
well after helium reionization has completed.

\section{Conclusions}
\label{sec:Conclusions}

Using a coupled radiative-transfer and Particle-Mesh $N$-body code, we
study the effects of helium reionization by a QSO source on the
surrounding IGM. We consider QSO source turn-on redshifts ranging from
$z=3.5$ to $z=5.5$, with the QSO spectrum modelled as a power law. In
the overdense environment expected for a QSO, the crossing of the
\HeIII\ ionization front boosts the temperature behind the front to
median values of $T\approx20-30\times10^3$~K, gradually relaxing to
$T\approx10\times10^3$~K by $z=2$ as thermal equilibrium between
photoionization heating and atomic radiative cooling processes and
adiabatic expansion cooling is established. The peak temperature tends
to be higher for earlier turn-on redshifts. A ``thermal proximity
effect'' results with a heightened IGM temperature within the \HeIII\
region produced. An identical source placed in an underdense
environment shows similar behaviour.

The median gas temperature surrounding the QSO, however, is not
uniform; it shows an increasing trend towards the \HeIII\ ionization
front, where spectral filtering by the intervening gas still hardens
the radiation field from the source in patches and where helium was
most recently ionized. We generate synthetic spectra to assess the
possibility of detecting the trend in the absorption lines along the
lines of sight to background neighbouring QSOs. The spectral signature
of the post-helium ionization temperature trends nearly vanishes in
the \HI\ Doppler parameters, which show both little evolution and
little dependence on distance from the source. For a source placed in
an overdense region, a median Doppler parameter of $25\kms$ at $z=3$
is found, independent of impact parameter and the time of \HeII\
reionization. A somewhat lower median Doppler parameter of $23\kms$ is
found for a source placed in the centre of a void. The median values
are in good agreement with the measured median Doppler parameter. This
is in contrast to simulations without radiative transfer, for which
the value is underpredicted by a few to several $\kms$
\citep{MBM01}. Late \HeII\ reionization appears to account for the
higher values. The absence of a dependence of the median Doppler
parameter on redshift and impact parameter may be attributed to the
absence of any temperature trend with redshift and impact parameter of
the denser regions which dominate the \Lya\ absorption spectra. By
contrast, a trend {\it is} found prior to full helium reionization,
with an increase in the Doppler parameters for a smaller impact
parameter of the line of sight, reflecting similar behaviour in the
peculiar velocity field. Thus the disappearance of a trend with impact
parameter may indicate the onset of full helium reionization.

A near polytropic relation is obtained between the gas temperature and
density with a slope that is close to adiabatic, extending to
overdensities of $\sim10^2$ in the vicinity of the source placed in a
massive halo. The temperatures found, approaching $10^5$~K, are those
expected for accretion onto halos and sheets, with radiative cooling
lowering the temperature at higher overdensities. In regions without a
massive halo, the turnover in the temperature occurs at overdensities
of only a few, again as expected from radiative cooling losses. No
inverted temperature--density relation is found in the underdense
gas. Whilst the temperature--density relation flattens soon after
\HeII\ reionization, the temperature in the underdense gas follows a
general trend of increasing with density, although with a wide spread
in values.

The detection of metal absorption systems would allow a more direct
probe of the gas temperature. Because the peculiar velocity
contribution to the velocity width of an absorption line would be
identical for features arising from different elements, while the
thermal contribution (in quadrature) decreases like the mass of the
ion, it is possible to separate these two contributions provided
absorption features due to two or more elements from the same
structure are detected. Such temperatures have been measured for a few
systems, although the corresponding \HI\ column densities tend to be
high, typically exceeding $10^{14}\,{\rm cm^{-2}}$, corresponding to
overdensities of a few to several (eg, \cite{2007arXiv0711.3358M}).
The temperature trend as a function of impact parameter is less
clearly defined for such overdensities, with a weak trend of higher
values for smaller impact parameters. None the less, a spread with
impact parameter could indicate the onset of full helium reionization,
particularly if very high temperatures exceeding $30\times10^3$~K are
found at $z<4$.

It would be very interesting to use metal absorption features arising
in the very diffuse IGM, corresponding to \HI\ column densities of
$10^{13}\,{\rm cm^{-2}}$ and lower, to probe the temperature structure
of the IGM, both before and after full helium reionization. There are
currently no detections of individual metal absorption features in
such low density diffuse systems, but neither do observations preclude
them. Exploiting them for temperature measurements, however, may need
to await spectroscopy on a much larger future generation of telescopes
such as the Thirty Meter Telescope or an Extremely Large Telescope.

\section*{Acknowledgements}
The computations reported here were performed using the SUPA
Astrophysical HPC facility and a facility funded by an STFC
Rolling-Grant. E.T. is supported by an STFC Rolling-Grant. C.B. was
supported by a grant from the Robert Cormack Bequest.

\appendix

\section{Radiative transfer on a polar grid}
\label{sec:RTgrid}

The equation of radiative transfer is solved along radii emanating
from the source using a photon-conserving probablistic method
\citep{1999ApJ...523...66A, BMW04, 2007MNRAS.380.1369T}. The radiative
transfer was performed on a polar grid with the start of each line of
sight placed at the source. The polar grid is divided into sectors
bounded by intervals in radius and angle. Since the optical depth must
be computed from the source to each polar cell, a density field must
be defined on the polar grid in order to compute the column densities
along the radii from the source. We considered three methods to
generate a smooth density field matching that of the Cartesian
grid. We describe the final method adopted first, but briefly describe
two alternative methods as well that were found inadequate for
information only should any of these methods be attempted.

In the method used, the PM mass distribution is apportioned to the
cells in the polar grid by dividing each particle into 500
sub-particles, spread over a comoving radius of $0.25\,{\rm
  h^{-1}}$~Mpc, and assigning the contributions to the polar grid
cells in which each sub-particle lies. The sub-particle density
distribution is cone-shaped, declining like $1/r$ in 2D, so that the
number per ring of fixed width $dr$ is constant. (The corresponding
distribution in 3D would decline like $1/r^2$ for an equal number per
shell.) The mass density on the polar grid is then computed as the
total mass in each cell divided by the cell volume, which increases
with radius. The method was found to produce a smooth rendition of the
density field without discontinuities and in good agreement with the
original density field measured on a Cartesian grid.

We considered two other methods that, though seemingly
straightforward, produced undesirable artefacts. In the first, the
density field defined on the Cartesian grid was re-mapped onto the
polar grid by computing the total mass contributed by the Cartesian
grid cells that overlapped a given polar cell. Because of the
irregular shapes of the overlap regions, it was found impractical to
compute the regions of overlap directly. We instead computed the
overlap regions using Monte Carlo integration by populating the
Cartesian grid uniformly with a large number of randomly distributed
points in order to compute the fraction of overlap with individual
polar cells. Each Cartesian grid cell was then assigned a weight for
each polar cell according to the amount of overlap. Once computed, the
weights were stored, as they need not be computed again for a given
source placement. The method, however, was found to produce
discontinuities in the density field if the source is placed near a
Cartesian cell boundary. The resulting density field on the polar grid
was also found to be excessively smoothed compared with the density
field defined on a Cartesian grid. A second approach was to compute
the density field directly on the polar grid using the N-body
particles. A nearest grid point method based on counting the particles
in individual polar cells fails, particularly near the centre of the
grid, because the cells are so small that large Poisson fluctuations
in the density field are produced, including empty cells with zero
density. Enlargening the cells near the centre results in poor
resolution and over-smoothing.

\bibliographystyle{mn2e-eprint}
\bibliography{apj-jour,biblio}
\label{lastpage}

\end{document}